\documentclass[
    ,final            
    ,numberedheadings 
  ]
  {aipproc}

\layoutstyle{6x9}

\usepackage{bm,amsfonts,amsmath,amssymb} 

\allowdisplaybreaks

\begin{document}

\title{Gravitational Radiation from Two-Body Systems\,\footnote{To
appear in the Proceedings of the Spanish Relativity Meeting ``A Century
of Relativity Physics'' (ERE05), Edited by Lysiane Mornas and Joaquin
Diaz-Alonso.}}


\author{Luc Blanchet} {address={Gravitation et Cosmologie
       (${\mathcal{G}}{\mathbb{R}}
       \varepsilon{\mathbb{C}}{\mathcal{O}}$), Institut d'Astrophysique
       de Paris, \\98$^{\text{bis}}$ boulevard Arago, 75014 Paris,
       France}}

\begin{abstract}
Thanks to the new generation of gravitational wave detectors LIGO and
VIRGO, the theory of general relativity will face new and important
confrontations to observational data with unprecedented precision.
Indeed the detection and analysis of the gravitational waves from
compact binary star systems requires beforehand a very precise solution
of the two-body problem within general relativity. The approximation
currently used to solve this problem is the post-Newtonian one, and must
be pushed to high order in order to describe with sufficient accuracy
(given the sensitivity of the detectors) the inspiral phase of compact
bodies, which immediately precedes their final merger. The resulting
post-Newtonian ``templates'' are currently known to 3.5PN order, and are
used for searching and deciphering the gravitational wave signals in
VIRGO and LIGO.
\end{abstract}

\maketitle

\section{Introduction}\label{secI}

A compelling motivation for accurate computations of the gravitational
radiation field generated by compact binary systems (\textit{i.e.}, made
of neutron stars and/or black holes) is the need for accurate
\textit{templates} to be used in the data analysis of the current and
future generations of laser interferometric gravitational wave
detectors. It is indeed recognized that the \textit{inspiral} phase of
the coalescence of two compact objects represents an extremely important
source for the ground-based detectors such as LIGO and VIRGO, provided
that their total mass does not exceed say 10 or 20 $M_\odot$ (this
includes the very interesting case of double neutron-star systems), and
for space-based detectors like LISA, in the case of the coalescence of
two galactic black holes, if the masses are within the range between say
$10^5$ and $10^8\,M_\odot$.

For these sources the \textit{post-Newtonian} (PN) approximation scheme
has proved to be the appropriate theoretical tool in order to construct
the necessary templates. A program started long ago with the goal of
obtaining these templates with 3PN and even 3.5PN
accuracy.\,\footnote{Following the standard custom we use the qualifier
$n$PN for a term in the wave form or (for instance) the energy flux
which is of the order of $1/c^{2n}$ relatively to the lowest-order
Newtonian quadrupolar radiation.} Several studies,
\textit{e.g.}~\cite{3mn,CF94}, have shown that such a high PN precision
is probably sufficient, not only for detecting the signals in
LIGO/VIRGO, but also for analyzing them and accurately measuring the
parameters of the binary (such high-accuracy templates will also be of
great value for detecting massive black-hole mergers in LISA). The
templates have been first completed through 2PN order~\cite{BDIWW95}.
The 3.5PN accuracy (in the case where the compact objects have
negligible intrinsic spins) has been achieved more
recently~\cite{BFIJ02,BDEI04}.

The calculation of the 3PN order turned out to be very intricate and
quite subtle. The first step has been to compute all the terms, in both
the 3PN equations of motion~\cite{JaraS98,JaraS99,BF00,BFeom,ABF01} and
3.5PN gravitational radiation field~\cite{B98tail,BIJ02,BI04mult}, by
means of the Hadamard self-field
regularization~\cite{Hadamard,Schwartz}. A regularization is needed in
this problem in order to remove the infinite self-field of point masses.
However, a few terms were left undetermined by Hadamard's
regularization, which correspond to some incompleteness of this
regularization occurring at the 3PN order. These terms could be
parametrized by some unknown numerical coefficients called
\textit{ambiguity parameters}. The second step has been to use the more
powerful \textit{dimensional regularization}~\cite{tHooft}, which is
technically based on analytic continuation in the dimension of space,
which finally enabled to fix the values of all the ambiguity
parameters~\cite{DJSdim,BDE04,BDEI04,BDEI05dr}.

In Section~\ref{secII} of this article we review and comment on the
striking appearance of Hadamard self-field regularization parameters at
3PN order, and on their computation using dimensional regularization.
Section~\ref{secIII} is devoted to the notion of the multipole moments
of an isolated post-Newtonian extended source, at the basis of the
construction of gravitational-wave post-Newtonian templates. In
Section~\ref{secIV} we present two checks of the values of the latter
ambiguity parameters, coming from the comparison between the binary's
dipole moment and its center-of-mass vector on the one hand, and based
on an argument from classical field-theory diagrams on the other hand.
Finally, in Section~\ref{secV}, we consider the limiting case where one
of the masses is exactly zero, and the remaining one moves with uniform
velocity, and show that such ``boosted Schwarzschild solution'' limit
yields the determination of the third ambiguity parameter in the
radiation field. These tests, altogether, provide a verification,
independent of dimensional regularization, for all the ambiguity
parameters in the 3PN gravitational radiation field.

\section{Hadamard regularization parameters}\label{secII}

The standard Hadamard regularization yields some ambiguous results for
the computation of certain integrals at the 3PN order, as Jaranowski and
Sch\"afer~\cite{JaraS98,JaraS99} first noticed in their computation of
the equations of motion of point particles within the ADM-Hamiltonian
formulation of general relativity. Hadamard's regularization is based on
the notion of \textit{partie finie} of a singular function, given by the
angular integral of the finite part coefficient in the singular
expansion of that function near a singular point, and the related notion
of \textit{partie finie} of a divergent integral. It was
shown~\cite{JaraS98,JaraS99} that there are \textit{two and only two}
types of ambiguous terms in the 3PN Hamiltonian, which were then
parametrized by two unknown numerical coefficients
$\omega_\mathrm{static}$ and $\omega_\mathrm{kinetic}$.

Motivated by the previous result, Blanchet and Faye introduced an
extended version of Hadamard's regularization~\cite{BFreg, BFregM},
which is mathematically well-defined and free of ambiguities; in
particular it yields unique results for the computation of any of the
integrals occuring in the 3PN equations of motion. Unfortunately, the
extended Hadamard regularization turned out to be in a sense incomplete,
because it was found~\cite{BF00, BFeom} that the 3PN equations of motion
involve {\it one and only one} unknown numerical constant, called
$\lambda$, which cannot be determined within the method. The comparison
with the work~\cite{JaraS98, JaraS99}, on the basis of the computation
of the invariant energy of compact binaries moving on circular orbits,
revealed~\cite{BF00} that
\begin{eqnarray}
  \omega_\mathrm{kinetic} &=& \frac{41}{24}\,,
  \label{omkin}
  \\ \omega_\mathrm{static} &=& - \frac{11}{3} \lambda -
  \frac{1987}{840}\,.
  \label{omstat}
\end{eqnarray}%
Therefore, the ambiguity $\omega_\mathrm{kinetic}$ is fixed, while
$\lambda$ is equivalent to the other ambiguity $\omega_\mathrm{static}$.
Notice that the value~(\ref{omkin}) for the kinetic ambiguity parameter
$\omega_\mathrm{kinetic}$, which is in factor of some velocity dependent
terms, is the only one for which the 3PN equations of motion are
Poincar\'e invariant. Fixing up this value was possible because the
extended Hadamard regularization~\cite{BFreg,BFregM} was defined in such
a way that it keeps the Poincar\'e invariance.

The appearance of one and only one physical unknown coefficient
$\lambda$ in the equations of motion constitutes a quite striking fact,
that is related specifically with the use of some Hadamard-type
regularization. Technically speaking, the presence of the parameter
$\lambda$ is associated with the so-called ``non-distributivity'' of
Hadamard's regularization.\,\footnote{By non-distributivity we mean that
the Hadamard regularization of a product of functions differs in general
from the product of regularizations.} Mathematically speaking, $\lambda$
is probably related to the fact that it is impossible to construct a
distributional derivative operator satisfying the Leibniz rule for the
derivation of the product. The Einstein field equations can be written
into many different forms, by shifting the derivatives and operating
some terms by parts with the help of the Leibniz rule. All these forms
are equivalent in the case of regular sources, but since the
distributional derivative operator violates the Leibniz rule they become
inequivalent for point particles. Finally, physically speaking, we can
argue that $\lambda$ has its root in the fact that, in a complete
computation of the equations of motion valid for two regular
\textit{extended} weakly self-gravitating bodies, many non-linear
integrals, when taken \textit{individually}, start depending, from the
3PN order, on the internal structure of the bodies, even in the
``compact-body'' limit where the radii tend to zero. However, when
considering the full equations of motion, we expect that all the terms
depending on the internal structure can be removed, in the compact-body
limit, by a coordinate transformation (or by some appropriate shifts of
the central world lines of the bodies), and that finally $\lambda$ is
given by a pure number, for instance a rational fraction, independent of
the details of the internal structure of the compact bodies. From this
argument (which could be justified by invoking the effacing principle in
general relativity~\cite{D83houches}) the value of $\lambda$ is
necessarily the one we shall obtain below, Eq.~(\ref{lambda}), and will
be valid for any compact objects, for instance black holes.
 
The ambiguity parameter $\omega_\mathrm{static}$, which is in factor of
some static, velocity-independent term, was computed by Damour,
Jaranowski and Sch\"afer~\cite{DJSdim} by means of \textit{dimensional
regularization}, instead of some Hadamard-type one, within the
ADM-Hamiltonian formalism. Their result is
\begin{equation}\label{omegas}
  \omega_\mathrm{static}=0\,.
\end{equation}
As Damour \textit{et al.}~\cite{DJSdim} argue, clearing up the static
ambiguity is made possible by the fact that dimensional regularization,
contrary to Hadamard's regularization, respects all the basic properties
of the algebraic and differential calculus of ordinary functions:
associativity, commutativity and distributivity of point-wise addition
and multiplication, Leibniz's rule, and the Schwarz lemma. In this
respect, dimensional regularization is certainly better than Hadamard's
one, which does not respect the distributivity of the product and
unavoidably violates at some stage the Leibniz rule for the
differentiation of a product.

The ambiguity parameter $\lambda$ is fixed from the result
(\ref{omegas}) and the necessary link~(\ref{omstat}) provided by the
equivalence between the harmonic-coordinates and ADM-Hamiltonian
formalisms. However, $\lambda$ was also computed directly by Blanchet,
Damour and Esposito-Far\`ese~\cite{BDE04} applying dimensional
regularization to the 3PN equations of motion in harmonic coordinates
(in the line of Refs.~\cite{BF00, BFeom}). The end result,
\begin{equation}\label{lambda}
  \lambda=-\frac{1987}{3080}\,,
\end{equation}
is in full agreement with Eq.~(\ref{omegas}). Besides the independent
confirmation of the value of $\omega_\mathrm{static}$ or $\lambda$, the
work~\cite{BDE04} provides also a confirmation of the
\textit{consistency} of dimensional regularization, because the explicit
calculations are entirely different from the ones of Ref.~\cite{DJSdim}:
harmonic coordinates are used instead of ADM-type ones, the work is at
the level of the equations of motion instead of the Hamiltonian, a
different form of Einstein's field equations is solved by a different
iteration scheme.

Let us comment here that the use of a self-field regularization, be it
dimensional or based on Hadamard's partie finie, signals a somewhat
unsatisfactory situation on the physical point of view, because,
ideally, we would like to perform a complete calculation valid for
extended bodies, taking into account the details of the internal
structure of the bodies (energy density, pressure, internal velocity
field). By considering the limit where the radii of the objects tend to
zero, one should recover the same result as obtained by means of the
point-mass regularization. This would demonstrate the suitability of the
regularization. This program was undertaken at the 2PN order by
Grishchuk and Kopeikin~\cite{Kop85,GKop86} who derived the equations of
motion of two extended fluid balls, and obtained equations of motion
depending only on the two masses $m_1$ and $m_2$ of the compact bodies.
At the 3PN order we expect that the extended-body program should give
the value of the regularization parameter $\lambda$ (maybe after some
gauge transformation to remove the terms depending on the internal
structure). Ideally, its value should be confirmed by independent and
more physical methods. One such method is the one of Itoh and
Futamase~\cite{itoh1,itoh2}, who derived the 3PN equations of motion in
harmonic coordinates by means of a particular variant of the famous
``surface-integral'' method introduced long ago by Einstein, Infeld and
Hoffmann~\cite{EIH}. This approach is interesting because it is based on
the physical notion of extended compact bodies in general relativity,
and is free of the problems of ambiguities due to Hadamard's self-field
regularization. The end result of Refs.~\cite{itoh1,itoh2} is in
agreement with the complete 3PN equations of motion in harmonic
coordinates~\cite{BF00,BFeom} and, moreover, is unambiguous, as it does
determine the ambiguity parameter $\lambda$ to exactly the
value~(\ref{lambda}).

We next consider the problem of the binary's radiation field, where the
same phenomenon occurs, with the appearance of some Hadamard
regularization ambiguity parameters at 3PN order. More precisely,
Blanchet, Iyer and Joguet~\cite{BIJ02}, in their computation of the 3PN
compact binary's \textit{mass quadrupole moment} $\mathrm{I}_{ij}$,
found it necessary to introduce \textit{three} Hadamard regularization
constants $\xi$, $\kappa$ and $\zeta$, which are additional to (and
independent of) the equation-of-motion related constant $\lambda$. The
total gravitational-wave flux at 3PN order, in the case of circular
orbits, was found to depend on a single combination of the latter
constants, $\theta = \xi+2\kappa+\zeta$, and the binary's orbital phase,
for circular orbits, involves only the linear combination of $\theta$
and $\lambda$ given by $\hat{\theta} = \theta-7\lambda/3$, as shown
in~\cite{BFIJ02}.

Dimensional regularization (instead of Hadamard's) was applied in
Refs.~\cite{BDEI04,BDEI05dr} to the computation of the 3PN radiation
field of compact binaries, finally leading to the following unique
values for the ambiguity parameters
\begin{eqnarray}
  \xi &=& - \frac{9871}{9240}\,, \label{xi}\\ \kappa &=& 0\,,
  \label{kappa}\\ \zeta &=& - \frac{7}{33}\,.\label{zeta}
\end{eqnarray}
These values represent the end result of dimensional regularization.
However, we shall review in the present Article some alternative
calculations which provide some checks, independent of dimensional
regularization, for all the parameters~(\ref{xi})--(\ref{zeta}).

The result~(\ref{xi})--(\ref{zeta}) completes the problem of the general
relativistic prediction for the templates of inspiralling compact
binaries up to 3PN order (and actually up to 3.5PN order as the
corresponding tail terms have already been determined~\cite{B98tail}).
The relevant combination of the parameters entering the 3PN energy flux
in the case of circular orbits is now fixed to be
\begin{equation}
\theta\equiv\xi+2\kappa+\zeta=-\frac{11831}{9240}\,.
\label{theta}\end{equation}
The orbital phase of compact binaries, in the adiabatic inspiral regime
(\textit{i.e.}, evolving by radiation reaction), involves at 3PN order a
combination of parameters which is determined as
\begin{equation}
\hat{\theta}\equiv \theta-\frac{7}{3}\lambda=\frac{1039}{4620}\,.
\label{thetahat}\end{equation}
The fact that the numerical value of this parameter is quite small,
$\hat{\theta}\simeq 0.22489$, indicates that the 3PN (or, even better,
3.5PN) order should provide an excellent approximation for both the
on-line search and the subsequent off-line analysis of gravitational
wave signals from inspiralling compact binaries in the LIGO and VIRGO
detectors.

\section{The multipolar post-Newtonian formalism}\label{secIII}

\subsection{Multipole moments of a post-Newtonian extended source}\label{secIIIA}

The multipole moments of a post-Newtonian (PN) source, by which we mean
a source which is at once slowly moving, weakly stressed and weakly
self-gravitating, are crucial for the present gravitational wave
generation formalism. They are obtained in Ref.~\cite{B98mult} as
functionals of the PN expansion of the pseudo-stress energy tensor
$\tau^{\mu\nu}$ of the matter and gravitational fields in the
\textit{harmonic coordinate} system. The pseudo-tensor $\tau^{\mu\nu}$
has a non-compact support because of the contribution of the
gravitational field which extends up to infinity from the source. Let us
denote the formal PN expansion of the pseudo tensor by means of an
overbar, so that $\overline{\tau}^{\mu\nu}=\mathrm{PN}[\tau^{\mu\nu}]$.
The two types of multipole moments of the gravitating source, mass-type
$\mathrm{I}_L$ moments and current-type ones $\mathrm{J}_L$, are then
given by\,\footnote{Our notation is: $L\equiv i_1\cdots i_\ell$ for a
multi-index composed of $\ell$ multipolar indices $i_1, \cdots, i_\ell$;
$x_L\equiv x_{i_1}\cdots x_{i_\ell}$ for the product of $\ell$ spatial
vectors $x^i\equiv x_i$; and $\hat{x}_L\equiv \mathrm{STF}(x_{i_1}\cdots
x_{i_\ell})$ for the symmetric-trace-free (STF) part of that product,
also denoted by carets surrounding the indices, $x_{\langle
L\rangle}\equiv\hat{x}_L$.}
\begin{eqnarray}\label{IL} \mathrm{I}_L(t)&=&
\frac{1}{c^2}\,\mathop{\mathrm{FP}}_{B=0}\int d^3\mathbf{x}\,r^B \left\{
\hat{x}_L\,\Bigl(\mathop{\overline{\tau}}_{[\ell]}{}^{\!\!00}+
\mathop{\overline{\tau}}_{[\ell]}{}^{\!\!ii}\Bigr)\right. \nonumber\\
&&\qquad\qquad \left. -\frac{4(2\ell+1)}{c(\ell+1)(2\ell+3)}
\,\hat{x}_{iL}\,\mathop{\dot{\overline{\tau}}}_{[\ell+1]}{}^{\!\!\!\!i0}\right.
\nonumber\\ &&\qquad\qquad \left.
+\frac{2(2\ell+1)}{c^2(\ell+1)(\ell+2)(2\ell+5)} \,\hat{x}_{ijL}
\,\mathop{\ddot{\overline{\tau}}}_{[\ell+2]}{}^{\!\!\!\!ij}\right\}\,,\\\label{JL}
\mathrm{J}_L(t)&=&
\frac{1}{c}\,\mathop{\mathrm{FP}}_{B=0}\varepsilon_{ab\langle i_\ell}
\int d^3\mathbf{x}\,r^B \left\{ \hat{x}_{L-1\rangle a}
\mathop{\overline{\tau}}_{[\ell]}{}^{\!\!b0}\right. \nonumber\\
&&\qquad\qquad \left.
-\frac{2\ell+1}{c(\ell+2)(2\ell+3)}\,\hat{x}_{L-1\rangle ac}
\mathop{\dot{\overline{\tau}}}_{[\ell+1]}{}^{\!\!\!\!bc}\right\}\,.
\end{eqnarray}
Since Eqs.~(\ref{IL})--(\ref{JL}) are valid only in the sense of PN
expansions, the operational meaning of the underscript $[\ell]$
in~(\ref{IL})--(\ref{JL}) is actually that of an infinite PN series,
which is given by
\begin{eqnarray}
\mathop{\overline{\tau}}_{[\ell]}{}^{\!\!\mu\nu}(\mathbf{x}, t) &=&
\sum_{k=0}^{+\infty}\,\alpha_{k,\ell}
\,\left(\frac{r}{c}\frac{\partial}{\partial t}\right)^{2k}
\overline{\tau}^{\mu\nu}(\mathbf{x},t)\,,\label{PNseries}\\
\alpha_{k,\ell} &=&
\frac{(2\ell+1)!!}{(2k)!!(2\ell+2k+1)!!}\,.\label{alphakl}
\end{eqnarray}

A basic feature of the expressions of the moments is that the integral
formally extends over the whole support of the PN expansion of the
stress-energy pseudo-tensor, $\overline{\tau}^{\mu\nu}$, \textit{ i.e.}
from $r\equiv\vert\mathbf{x}\vert=0$ up to infinity. Recall that a
formal PN series such as $\overline{\tau}^{\mu\nu}$ is physically
meaningful only within the near-zone. Therefore the
integrals~(\ref{IL})--(\ref{JL}) physically refer to a result obtained
from near-zone quantities only (in the formal limit where $c \rightarrow
+\infty$). However, it was found extremely useful in Ref.~\cite{B98mult}
to mathematically extend the integrals up to $r\rightarrow +\infty$.
This was made possible by the use of the prefactor $r^B$, together with
a process of analytic continuation in the complex $B$
plane.\,\footnote{The prefactor $r^B$ should in principle be
adimensionalized as $(r/r_0)^B$ where $r_0$ is a constant arbitrary
scale, but here we set $r_0=1$.} This shows up in
Eqs.~(\ref{IL})--(\ref{JL}) as the crucial Finite Part (FP) operation,
when $B\rightarrow 0$, which technically allows one to uniquely define
integrals which would otherwise be divergent at their upper boundary,
$r=\vert\mathbf{x}\vert\rightarrow +\infty$. See Ref.~\cite{B98mult} for
the proof and details.

\subsection{Surface-integral expressions of the multipole moments}\label{secIIIB}

Let us next review the recent derivation~\cite{BDI04zeta} of an
alternative form of the PN source moments~(\ref{IL})--(\ref{JL}) in
terms of two-dimensional surface integrals. Such a possibility of
expressing the moments, for general $\ell$ and at any PN order, as some
surface integrals is quite useful for practical purposes, as we shall
show in the application we consider in Section~\ref{secV}. In keeping
with the fact that the ``volume integrals'' Eqs.~(\ref{IL})--(\ref{JL})
physically involve only near-zone quantities, the ``surface integrals''
into which we shall transform the moments $\mathrm{I}_L$ and
$\mathrm{J}_L$ physically refer to an operation which extracts some
coefficients in the ``far near-zone'' expansion of the gravitational
field, \textit{i.e.} in the expansion in increasing powers of $1/r$ of
the PN-expanded near-zone metric. Technically, as our starting
point~(\ref{IL})--(\ref{JL}) is made of integrals extended up to
$r\rightarrow +\infty$, our mathematical manipulations below will
involve ``surface terms'' on arbitrary large spheres $r = \mathcal{R}$.
All these manipulations will be mathematically well-defined because of
the properties of complex analytic continuation in $B$.

The basic idea is to go from the ``source term'',
$\overline{\tau}^{\mu\nu}$, to the corresponding ``solution''
$\overline{h}^{\mu\nu}$, \textit{via} integrating by parts the Laplace
operator present in the Einstein field equation in harmonic coordinates,
namely $\overline{\tau}^{\mu\nu}=\frac{c^4}{16\pi
G}\,\square\,\overline{h}^{\mu\nu}$, where $\overline{h}^{\mu\nu}$ is
the (PN expansion of the) basic gravitational field variable, satisfying
the harmonic-coordinate condition $\partial_\nu\overline{h}^{\mu\nu}=0$.
From Eq.~(\ref{PNseries}) we have
\begin{equation}\label{PNseries2}
\int d^3\mathbf{x}\,r^B
\,\hat{x}_L\,\mathop{\overline{\tau}}_{[\ell]}{}^{\!\!\mu\nu} =
\frac{c^4}{16\pi G}\sum_{k=0}^{+\infty}\,\alpha_{k,\ell}
\,\left(\frac{d}{c dt}\right)^{\!2k}\int
d^3\mathbf{x}\,r^{B+2k}\,\hat{x}_L\,\Box\,\overline{h}^{\mu\nu}\,,
\end{equation}
in the right-hand-side of which we insert
$\Box=\Delta-\left(\frac{\partial}{c\,\partial t}\right)^2$, and operate
the Laplacian by parts using
$\Delta(r^{B+2k}\,\hat{x}_L)=(B+2k)(B+2\ell+2k+1)r^{B+2k-2}\,\hat{x}_L$.
In the process we can ignore the all-integrated surface terms because
they are identically zero by complex analytic continuation, from the
case where the real part of $B$ is chosen to be a large enough
\textit{negative} number. Using the expression of the coefficients
(\ref{alphakl}), we are next led to the alternative expression
\begin{equation}\label{PNseries3}
\int d^3\mathbf{x}\,r^B
  \,\hat{x}_L\,\mathop{\overline{\tau}}_{[\ell]}{}^{\!\!\mu\nu} =
  \frac{c^4}{16\pi
  G}\sum_{k=0}^{+\infty}\,B(B+2\ell+4k+1)\,\alpha_{k,\ell}
  \left(\frac{d}{cdt}\right)^{\!2k}\!\!\int
  d^3\mathbf{x}\,r^{B+2k-2}\,\hat{x}_L\,\overline{h}^{\mu\nu}\,.
\end{equation}
A remarkable feature of this result, which is the basis of our new
expressions, is the presence of an \textit{explicit factor} $B$ in front
of the integral. The factor means that the result depends only on the
occurrence of \textit{poles}, $\propto 1/B^p$, in the boundary of the
integral at infinity: $r\rightarrow +\infty$ with $t=\mathrm{const}$.

Thanks to the factor $B$ we can replace the integration domain of
Eq.~(\ref{PNseries3}) by some outer domain of the type $r>\mathcal{R}$,
where $\mathcal{R}$ denotes some large arbitrary constant radius. The
integral over the inner domain $r<\mathcal{R}$ is always zero in the
limit $B\rightarrow 0$ because the integrand is constructed from
$\overline{\tau}^{\mu\nu}$, and we are considering extended regular PN
sources, without singularities. Now, in the outer (but still near-zone)
domain we can replace the PN metric coefficients $\overline{h}^{\mu\nu}$
by the expansion in increasing powers of $1/r$ of the PN-expanded
metric, which is identical to the multipolar expansion of the
PN-expanded metric, that we shall denote by
$\mathcal{M}\bigl(\overline{h}^{\mu\nu}\bigr)$. Hence we have
\begin{equation}\label{PNseries4}
\int d^3\mathbf{x}\,r^B
\,\hat{x}_L\mathop{\overline{\tau}}_{[\ell]}{}^{\!\!\mu\nu} =
\frac{c^4}{16\pi G}\sum_{k=0}^{+\infty}B(B+2\ell+4k+1)\,\alpha_{k,\ell}
\left(\frac{d}{cdt}\right)^{\!2k}\!\!\!\int_{r>\mathcal{R}}\!\!
d^3\mathbf{x}\,r^{B+2k-2}\hat{x}_L\mathcal{M}\bigl(
\overline{h}^{\mu\nu}\bigr)\,.
\end{equation}
We want now to make use of a more explicit form of the far near-zone
expansion $\mathcal{M}\bigl(\overline{h}^{\mu\nu}\bigr)$, whose general
structure is known. It consists of terms proportional to arbitrary
powers of $1/r$, and multiplied by powers of the \textit{logarithm} of
$r$; more precisely,
\begin{equation}\label{Mhexp}
\mathcal{M}\bigl(\overline{h}^{\mu\nu}\bigr)(\mathbf{x},t)=\sum_{a,\,b}
\frac{(\ln r)^b}{r^a}\,\varphi_{a,b}^{\mu\nu}(\mathbf{n},t)\,,
\end{equation}
where $a$ can take any positive or negative integer values, and $b$ can
be any positive integer: $a\in\mathbb{Z}$, $b\in\mathbb{N}$. The
coefficients $\varphi_{a,b}^{\mu\nu}$ depend on the unit direction
$\mathbf{n}\equiv\mathbf{x}/r$ and on the coordinate time $t$ (in the
harmonic coordinate system). The structure~(\ref{Mhexp}) for the
multipolar expansion of the near-zone (PN-expanded) metric is a
consequence of the so-called matching equation
\begin{equation}\label{match}
\mathcal{M}\bigl(\overline{h}^{\mu\nu}\bigr)\equiv
\overline{\mathcal{M}\left(h^{\mu\nu}\right)}\,,
\end{equation}
which says that the multipolar re-expansion of the PN metric
$\overline{h}^{\mu\nu}$ agrees, in the sense of formal series, with the
\textit{near-zone} re-expansion (also denoted with an overbar) of the
external multipolar metric $\mathcal{M}\left(h^{\mu\nu}\right)$
(see~\cite{B98mult} for details). Inserting Eq.~(\ref{Mhexp}) into
(\ref{PNseries4}), we are therefore led to the computation of the
integral
\begin{equation}\label{intexp}
\int_{r>\mathcal{R}}\!\!d^3\mathbf{x}\,r^{B+2k-2}\,\hat{x}_L
\,\mathcal{M}\bigl(\overline{h}^{\mu\nu}\bigr)=
\sum_{a,\,b}\int_\mathcal{R}^{+\infty}dr\,r^{B+2k+\ell-a}\,(\ln
r)^b\int d\Omega\,\hat{n}_L\,\varphi_{a,b}^{\mu\nu}(\mathbf{n},t)\,,
\end{equation}
where $d\Omega$ is the solid angle element associated with the unit
direction $\mathbf{n}$ (and $\hat{n}_L\equiv \hat{x}_L/r^\ell$). The
radial integral can be trivially integrated by analytic continuation in
$B$, with result
\begin{equation}\label{radial}
\int_\mathcal{R}^{+\infty}dr\,r^{B+2k+\ell-a}\,(\ln
r)^b=-\left(\frac{d}{dB}\right)^b\left[\frac{
\mathcal{R}^{B+2k+\ell-a+1}}{B+2k+\ell-a+1}\right]\,.
\end{equation}
Remember that we are ultimately interested only in the analytic
continuation of such integrals down to $B=0$. And as an integral such
as~(\ref{radial}) is multiplied by a coefficient which is proportional
to $B$, we must control the poles of Eq.~(\ref{radial}) at $B=0$. Those
poles are in general multiple because of the presence of powers of $\ln
r$ in the expansion, and the consecutive multiple differentiation with
respect to $B$ shown in Eq.~(\ref{radial}). The poles at $B=0$ clearly
come from a single value of $a$, namely $a=2k+\ell+1$. For that value,
the ``multiplicity'' of the pole takes the value $b + 1$. Here a useful
simplification comes from the fact that the factor in front of the
integrals in~(\ref{PNseries4}) is of the form $\sim B(B+K)$. In other
words, this factor contains only the first and second powers of $B$.
Therefore, only the simple and double poles $1/B$ and $1/B^2$
in~(\ref{radial}) can contribute to the final result. Hence, we conclude
that it is enough to consider the values $b=0,1$ for the exponent $b$ of
$\ln r$ in the expansion~(\ref{Mhexp}).

To express the result in the most convenient manner let us introduce a
special notation for some relevant combination of coefficients
$\varphi_{a,b}^{\mu\nu}(\mathbf{n},t)$, which as we just said correspond
exclusively to the values $a=\ell+2k+1$ and $b=0$ or $1$. Namely,
\begin{equation}\label{Psi}
\Psi_{k,\ell}^{\mu\nu}(\mathbf{n},t)\equiv
\alpha_{k,\ell}\Bigl[-(2\ell+4k+1)\varphi_{2k+\ell+1,
0}^{\mu\nu}(\mathbf{n},t) + \varphi_{2k+\ell+1,1}^{\mu\nu}(\mathbf{n},t)\Bigr]\,,
\end{equation}
in which we have absorbed the numerical coefficient $\alpha_{k,\ell}$
defined by~(\ref{alphakl}). With this notation we then obtain
\begin{equation}\label{intexpres}
\mathop{\mathrm{FP}}_{B=0}\,B\,(B+2\ell+4k+1)\,\alpha_{k,\ell}
\int_{r>\mathcal{R}}\!\!d^3\mathbf{x}\,r^{B+2k-2}\,\hat{x}_L
\,\mathcal{M}\bigl(\overline{h}^{\mu\nu}\bigr)=4\pi
\left\langle\,\hat{n}_L\,\Psi_{k,\ell}^{\mu\nu}\right\rangle\,,
\end{equation}
where the brackets refer to the spherical or angular average (at
coordinate time $t$), \textit{i.e.}
\begin{equation}\label{angle}
\left\langle\,\hat{n}_L\,\Psi_{k,\ell}^{\mu\nu}\right\rangle (t)
\equiv\int
\frac{d\Omega}{4\pi}\,\hat{n}_L\,\Psi_{k,\ell}^{\mu\nu}(\mathbf{n},t)\,.
\end{equation}
The quantities~(\ref{angle}) are integrals over a unit sphere, and can
rightly be referred to as \textit{surface integrals}. These surface
integrals are the basic blocks entering our alternative expressions for
the multipole moments. If we wish to physically think of them as
integrals over some two-surface surrounding the source, we can roughly
consider that this two-surface is located at a radius $\mathcal{R}$,
with $a \ll \mathcal{R} \ll c\,T$. Anyway, the important point is that,
as we can see from Eq.~(\ref{angle}), the surface integrals, and
therefore the multipole moments, are strictly independent of the choice
of the intermediate scale $\mathcal{R}$ which entered our reasoning.

Finally, we are in a position to write down the following final results
for an alternative form of the source multipole
moments~(\ref{IL})--(\ref{JL}), expressed solely in terms of the surface
integrals of the type~(\ref{angle}),
\begin{eqnarray}
\mathrm{I}_L &=& \frac{c^2}{4\,G}\,\sum_{k=0}^{+\infty}\biggl\{
\left(\frac{d}{cdt}\right)^{\!\!2k} \left\langle\,\hat{n}_L
\left(\Psi_{k,\ell}^{00}+\Psi_{k,\ell}^{ii}\right)\right\rangle\nonumber\\
&&\qquad\qquad -\frac{4(2\ell+1)}{(\ell+1)(2\ell+3)}
\,\left(\frac{d}{cdt}\right)^{\!\!2k+1} \left\langle\,\hat{n}_{iL}
\,\Psi_{k,\ell+1}^{i0}\right\rangle\nonumber\\ &&\qquad\qquad
+\frac{2(2\ell+1)}{(\ell+1)(\ell+2)(2\ell+5)}\,
\left(\frac{d}{cdt}\right)^{\!\!2k+2} \left\langle\,\hat{n}_{ijL}
\,\Psi_{k,\ell+2}^{ij}\right\rangle\biggr\}\,,\label{ILfinal}\\
\mathrm{J}_L &=& \frac{c^3}{4\,G}\,\varepsilon_{ab\langle
i_\ell}\sum_{k=0}^{+\infty}\biggl\{ \left(\frac{d}{cdt}\right)^{\!\!2k}
\left\langle\,\hat{n}_{L-1\rangle
a}\,\Psi_{k,\ell}^{b0}\right\rangle\nonumber\\ &&\qquad\qquad
-\frac{2\ell+1}{(\ell+2)(2\ell+3)}
\,\left(\frac{d}{cdt}\right)^{\!\!2k+1}
\left\langle\,\hat{n}_{L-1\rangle ac}
\,\Psi_{k,\ell+1}^{bc}\right\rangle\biggr\}\label{JLfinal}\,.
\end{eqnarray}

\section{Multipole moments of two-body systems}\label{secIV}

\subsection{Quadrupole and dipole moments, and the center-of-mass vector}\label{secIVA}

Let us show how a particular combination of ambiguity parameters can be
determined within Hadamard's regularization and confirm the result of
dimensional regularization. For this purpose we use the computations in
Ref.~\cite{BI04mult} of the mass-type quadrupole $\mathrm{I}_{ij}$ and
dipole $\mathrm{I}_{i}$ moments of point particle binaries at the 3PN
order. These were derived by applying the expression~(\ref{IL}) [with
$\ell=1,2$] to a binary systems of point masses, following the rules of
the Hadamard regularization, in the so-called ``pure Hadamard-Schwartz''
(pHS) variant of it. Following the definition of Ref.~\cite{BDE04}, the
pHS regularization is a specific, minimal Hadamard-type regularization
of integrals, based on the usual Hadamard partie finie of a divergent
integral, together with a minimal treatment (supposed to be
``distributive'') of compact-support terms. The pHS regularization also
assumes the use of standard Schwartz distributional
derivatives~\cite{Schwartz}.

We shall denote by $\mathrm{I}^\mathrm{\,pHS}_{ij}$ the result of such
pHS calculation of the mass-type quadrupole moment. Now it was argued in
Ref.~\cite{BIJ02} that the Hadamard regularization of the 3PN quadrupole
moment is incomplete, in the sense that the pHS calculation
$\mathrm{I}^\mathrm{\,pHS}_{ij}$ must be augmented, in order to be
correct, by some unknown, ambiguous, contributions. The first source of
ambiguity is the ``kinetic'' one, linked to the inability of the
Hadamard regularization to ensure the global Poincar\'e invariance of
the formalism. As discussed in Ref.~\cite{BIJ02} (see also
Section~\ref{secII}) we must account for the kinetic ambiguity by adding
``by hands'' a specific ambiguity term, depending on a single ambiguity
parameter called $\zeta$. The second source of ambiguity is ``static''.
It comes from the \textit{a priori} unknown relation between some
Hadamard regularization length scales, $s_1$ and $s_2$ (one for each
particles), and the ones, called $r'_1$ and $r'_2$, parametrizing the
final 3PN equations of motion in harmonic coordinates~\cite{BF00,BFeom}.
The static ambiguity is accounted for by two other ambiguity parameters
$\xi$ and $\kappa$ (see Section~\ref{secII}).

The Hadamard-regularized 3PN quadrupole moment reads
\begin{eqnarray}\label{Iij}
\mathrm{I}_{ij}[\xi,\kappa,\zeta] &=& \mathrm{I}^\mathrm{\,pHS}_{ij}
\\&+& \frac{44}{3} \frac{G^2\,m_1^3}{c^6}\left[\left(\xi + \frac{1}{22}+
\kappa\frac{m_1+m_2}{m_1}\right)y_1^{\langle i}a_1^{j\rangle}
+\left(\zeta + \frac{9}{110}\right)v_1^{\langle ij\rangle}\right] +
1\leftrightarrow 2\,,\nonumber
\end{eqnarray}
where one sees in the second term the effect of adding the ambiguities,
parametrized by the same parameters $\xi$, $\kappa$ and $\zeta$ as
introduced in Ref.~\cite{BIJ02}. Here, $m_1$ and $m_2$ are the masses,
$y_1^i$, $v_1^i$ and $a_1^i$ denote the position, velocity and Newtonian
acceleration of the first particle, and we pose $y_1^{\langle
i}a_1^{j\rangle}\equiv\mathrm{STF}(y_1^{i}a_1^{j})$ and $v_1^{\langle
ij\rangle}\equiv\mathrm{STF}(v_1^{i}v_1^{j})$. The symbol
$1\leftrightarrow 2$ refers to the same terms but concerning the second
particles. All the terms composing the pHS part have been explicitly
computed up to 3PN order for general binary orbits~\cite{BI04mult}.

Let us now consider the case of the mass dipole moment $\mathrm{I}_{i}$.
Repeating the same arguments as for the quadrupole, we can write
$\mathrm{I}_{i}$ as the pHS part $\mathrm{I}^\mathrm{\,pHS}_i$ and
augmented by an ambiguous part. However, in the dipole case we find that
no ambiguity of the kinetic type occurs, and that the only ambiguity is
static. We find that the expression analogous to~(\ref{Iij}) reads
\begin{equation}\label{Ii}
\mathrm{I}_i[\xi+\kappa] = \mathrm{I}^\mathrm{\,pHS}_i + \frac{22}{3}
\,\frac{G^2\,m_1^3}{c^6}\left(\xi + \kappa + \frac{1}{22}
\right)\,a_1^i+1\leftrightarrow 2\,.
\end{equation}
As we see, there is only one ambiguity parameter, in the form of the
\textit{sum} of $\xi$ and $\kappa$, where $\xi$ and $\kappa$ are exactly
the same as in the quadrupole moment~(\ref{Iij}). Let us now fix that
particular sum of ambiguity parameters.

The case of the dipole moment $\mathrm{I}_i$ is very interesting. Indeed
let us argue that $\mathrm{I}_i$, which represents the distribution of
positions of particles as weighted by their \textit{gravitational
masses} $m_\mathrm{g}$, must be \textit{identical} to the position of
the center of mass $\mathrm{G}_i$ of the system of particles
(\textit{per} unit of total mass), because the center of mass
$\mathrm{G}_i$ represents in fact the same quantity as the dipole
$\mathrm{I}_i$ but corresponding to the \textit{inertial masses}
$m_\mathrm{i}$ of the particles. The equality between mass dipole
$\mathrm{I}_i$ and center-of-mass position $\mathrm{G}_i$ can thus be
seen as a consequence of the equivalence principle
$m_\mathrm{i}=m_\mathrm{g}$, which is surely incorporated in our model
of point particles. Now the center of mass $\mathrm{G}_i$ is already
known at the 3PN order for point particle binaries, as one of the
conserved integrals of the 3PN motion in harmonic
coordinates.\,\footnote{We neglect the radiation-reaction term at 2.5PN
order.} The point is that $\mathrm{G}_i$, given in Ref.~\cite{ABF01}, is
free of ambiguities; for instance the ambiguity parameter $\lambda$ in
the 3PN equations of motion disappears from the expression of
$\mathrm{G}_i$. Let us therefore impose the equivalence between
$\mathrm{I}_i$ and $\mathrm{G}_i$, which means that we make the complete
identification
\begin{equation}\label{IiGi}
\mathrm{I}_i[\xi+\kappa] \equiv \mathrm{G}_i\,.
\end{equation}
Comparing $\mathrm{I}_i$ with the expression of $\mathrm{G}_i$ given by
Eq.~(4.5) in~\cite{ABF01}, we find that Eq.~(\ref{IiGi}) is verified for
all the terms \textit{if and only if} the particular combination of
ambiguity parameters $\xi+\kappa$ takes the unique value
\begin{equation}\label{xikappa}
\xi + \kappa = -\frac{9871}{9240}\,.
\end{equation}
This result, obtained within Hadamard's regularization, is nicely
consistent with the result of dimensional regularization, see
Eqs.~(\ref{xi})--(\ref{kappa}). It shows that, although as we have seen
Hadamard's regularization is physically incomplete (at 3PN order), it
can nevertheless be partially completed by invoking some external
physical arguments --- in the present case the equivalence between mass
dipole and center-of-mass position. On the other hand, dimensional
regularization \textit{is} complete; it does not need to invoke any
external physical argument in order to determine the value of all the
ambiguity parameters. Nevertheless, it remains that the
result~(\ref{xikappa}), based simply on a consistency argument between
the 3PN equations of motion and the 3PN radiation field, does provide a
verification of the consistency and completeness of dimensional
regularization itself.

\subsection{Diagrammatic representation of the multipole moments}\label{secIVB}

Let us describe the multipole moments in terms of classical field-theory
diagrams, representing the non-linear interactions of classical general
relativity (we refer to~\cite{Dgef96} for definition and use of these
diagrams). We represent the basic delta-function sources entering the
matter stress-energy tensor $T^{\mu\nu}$ --- \textit{i.e.}, the matter
part of the pseudo-tensor $\overline{\tau}^{\mu\nu}$ of
Section~\ref{secIII} --- as two world-lines, and each (post-Minkowskian)
propagator $\Box^{-1}$ as a dotted line. The various non-linear
potentials entering the gravitational part of $\overline{\tau}^{\mu\nu}$
can then be represented by drawing some dotted lines which start at the
matter sources, join at some intermediate vertices, corresponding to
some non-linear couplings, and end at the field point $x$. Finally, we
can represent the inclusion of the multipolar factors, such as
$\hat{x}_L$, by adding a circled cross $\otimes$. It is then understood
that one integrates over the crossed vertex, \textit{i.e.}, the field
point.

\begin{figure}[h] \centering
   \includegraphics[width=6cm]{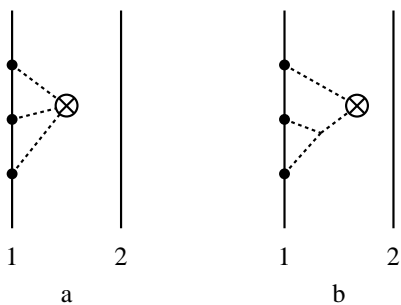}
 \caption {Dangerously divergent diagrams contributing to the 3PN
multipole moments. The world-lines of particles 1 and 2 are represented
by vertical solid lines, the propagator $\Box^{-1}$ by dotted lines, the
source points by bullets, and the $\otimes$ symbol means a
multiplication by a multipolar factor, such as $\hat{x}_L$, together
with a spatial integration $\int d^d \mathbf{x}$.}
   \label{fig}
\end{figure}

Using such a representation, the multipole moments are given by the sum
of many diagrams. We are now looking at ``dangerously'' diverging
diagrams, which generate poles $\propto 1/\varepsilon$ in a
dimensionally continued approach, with $d=3+\varepsilon$ being the
dimension of space. Examining the types of singular integrals
corresponding to the possible diagrams, we find~\cite{BDEI05dr} that the
only dangerously diverging diagrams are those containing (at least)
three propagator lines that can simultaneously shrink to zero size, as a
subset of vertices coalesce together on one of the particle world-lines.
But as there are, in the present problem dealing with the 3PN order, at
most three source points, this means that the dangerously divergent
diagrams are only those represented in Fig.~\ref{fig} (or their mirror
image obtained by exchanging $1 \leftrightarrow 2$).
\vspace{0.5cm}

Since the dangerous divergencies associated with the vicinity of the
first world-line (say) are entirely contained in the diagrams shown in
Fig.~\ref{fig}, they are, therefore, proportional to $m_1^3$
(\textit{i.e.}, one factor $m_1$ \textit{per} source point), without any
explicit\,\footnote{There is also an implicit dependence on $m_2$
\textit{via} the fact that the acceleration $a_1^i$ is proportional to
$m_2$. But, at the level of the diagrams, $a_1^i$ must be considered as
a pure characteristic of the first world-line.} dependence on the second
mass $m_2$. As a consequence, we can prove~\cite{BDEI05dr}, because the
presence of ambiguity parameters is directly linked with the occurence
of poles $\propto 1/\varepsilon$, that the structure of the ambiguous
terms in the mass quadrupole moment~(\ref{Iij}) must be such that it is
proportional to some factor $m_1^3$. Now, the definition of the
parameter $\kappa$ in Ref.~\cite{BIJ02} was to parametrize a conceivable
\textit{a priori} static ambiguity appearing in the renormalization of
the logarithmic divergencies of the quadrupole moment, and these
ambiguities were found in Eq.~(\ref{Iij}) to be of the form $\left(\xi +
\kappa\right) \, m_1^3 + \kappa \, m_1^2 \, m_2$ (for what concerns the
first particle). This shows that the parameter $\kappa$ corresponds to a
mixing between diagrams with three legs on the first world-line (as in
Fig.~\ref{fig}) and diagrams having two legs on the first world-line and
one on the second. Our diagrammatic study has shown that the latter
diagrams have no dangerous divergencies, \textit{i.e.}, that they do not
introduce any conceivable ambiguity. Therefore we conclude, confirming
Eq.~(\ref{kappa}), that
\begin{equation}
\kappa = 0\,.
\end{equation}

\section{Field generated by a single body}\label{secV}

As another application, making use of the explicit surface-integral
formula~(\ref{ILfinal}), and yielding another check of ambiguity
parameters, we wish to compute the source-type multipole moments of a
spherically symmetric extended body moving with \textit{uniform}
velocity. Remember that our formalism assumes, in principle, that we are
dealing with regular, weakly self-gravitating bodies. We expect, because
of the effacing properties of Einstein's theory~\cite{D83houches}, that
our final physical results, especially when they are expressed as
surface integrals like in~(\ref{ILfinal}), can be applied to more
general sources, such as neutron stars or black holes. Indeed, we are
going to confirm this expectation in the simplest possible case, that of
an isolated spherically symmetric body which is known, by Birkhoff's
theorem, to generate a universal exterior gravitational field, given by
the Schwarzschild solution.

\subsection{The boosted Schwarzschild solution}\label{secVA}

Following Ref.~\cite{BDI04zeta} we shall apply our formulas to a
\textit{boosted Schwarzschild solution} (BSS). Actually, in order to
justify our use of the BSS (in standard harmonic coordinates), we must
dispose of a small technicality. This technicality concerns the
non-uniqueness of harmonic coordinates for the Schwarzschild solution,
even under the assumption of stationarity (in the rest frame) and
spherical symmetry. Indeed, under these assumptions, and starting from
the usual Schwarzschild radial coordinate, say $r_S$, the (rest frame)
radial coordinate of the most general harmonic coordinate system, say $r
= k(r_S)$, must satisfy the differential equation (see, \textit{e.g.},
Weinberg~\cite{Weinberg}, page 181)
\begin{equation}\label{harm}
\frac{d}{d r_S}\left[\left(r_S^2 - \frac{2 \,G M}{c^2}\,r_S\right)
\frac{d k}{d r_S}\right] = 2\,k \,.
\end{equation}
The standard solution of Eq.~(\ref{harm}), which is considered in all
textbooks such as~\cite{Weinberg}, reads simply
\begin{equation}\label{standsol}
r = k^\mathrm{standard}(r_S) = r_S - \frac{G M}{c^2}\,.
\end{equation}
In the black hole case, the solution~(\ref{standsol}) is the only one
which is regular on the horizon, \textit{i.e.} when $r_S = 2 G M/c^2$.
However, in the case of the external metric of an extended spherically
symmetric body, regularity on the horizon is not a relevant issue. What
is relevant is that the solution of the \textit{external}
problem~(\ref{harm}) be smoothly matched to a \textit{regular} solution
of the corresponding \textit{internal} problem. As usual, this matching
determines a unique solution everywhere. In general, this unique,
everywhere regular, solution will correspond, in the exterior of the
body, to a particular case of the general, two-parameter solution of the
second-order differential equation~(\ref{harm}). The latter is of the
form
\begin{equation}\label{gensol}
r = k^\mathrm{general}(r_S) = C_1 \left(r_S - \frac{G M}{c^2} \right)+
C_2\,k_2(r_S) \,,
\end{equation}
where $k_2(r_S)$ denotes the (uniquely defined) ``radially decaying
solution'' of Eq.~(\ref{harm}), and where $C_1$ and $C_2$ are two
integration constants. Indeed, when considering the flat-space limit of
Eq.~(\ref{harm}), it is easily seen that there are two independent
solutions which behave, when $r_S \rightarrow +\infty$, as $r_S$ and
$r_S^{-2}$ respectively. An explicit expression for the decaying
solution is\,\footnote{Here $F(2,2,4,z)$ denotes a particular case of
Gauss' hypergeometric function
$$F(\alpha,\beta,\gamma,z)=1+\frac{\alpha\beta}{\gamma}\,
\frac{z}{1!}+\frac{\alpha(\alpha+1)
\beta(\beta+1)}{\gamma(\gamma+1)}\,\frac{z^2}{2!}+\cdots\,.$$}
\begin{equation}\label{decaysol}
k_2(r_S) = \frac{1}{r_S^2} \,F\left(2,2,4,\frac{2 \,G M}{c^2
r_S}\right)\,.
\end{equation}
We can always normalize $C_1$ to the value $C_1 = 1$. Then, with the
above definitions, $C_2$ has the dimension of a length cubed. By
considering in more detail the matching of the general solution of the
harmonically relaxed Einstein equations at the 2PN level (see,
\textit{e.g.}, the book by Fock~\cite{Fock}, page 322), one easily finds
that the second integration constant is of the order of $C_2 \sim (G
M/c^2)^2 \,a$, where $a$ denotes the radius of the extended body under
consideration. It is also easily checked that the constant $C_2$
parametrizes, at the linearized order, a \textit{gauge vector} of the
form $\varphi^i \propto C_2 \,\partial_i (1/r)$, and can thus be
referred to as a ``gauge parameter''.

Contrarily to the multipole moments of \textit{stationary} sources,
which are geometric invariants (and can be expressed as surface
integrals on a sphere at spatial infinity), the source multipole moments
defined in Ref.~\cite{B98mult} (and re-expressed in Section~\ref{secIII}
as surface integrals over spheres in some intermediate region, $ a \ll r
\ll c\,T$) are probably not geometric invariants. They are useful
intermediate constructs, which allow one to compute physically invariant
information, but their definition is linked to the choice of harmonic
coordinates covering the source. There are also various \textit{gauge
multipoles} (denoted $\mathrm{W}_L$, $\mathrm{X}_L$, $\mathrm{Y}_L$,
$\mathrm{Z}_L$ in Ref.~\cite{B98mult}) which will influence, at some
non-linear order, the values of the two sequences of \textit{physical
multipoles}: $\mathrm{I}_L$, $\mathrm{J}_L$. Therefore, one should
expect that, at some non-linear order, the physical multipoles
$\mathrm{I}_L$, $\mathrm{J}_L$ of a boosted general, harmonic-coordinate
spherically symmetric metric will start to depend on the value of the
gauge parameter $C_2$.

Here, we are only interested in computing the quadrupole moment
$\mathrm{I}_{ij}$ of a boosted general spherically symmetric metric. We
shall see below that the index structure of $\mathrm{I}_{ij}$ will be
provided by the STF tensor product of the boost velocity $V^i$ with
itself, denoted $V^{\langle i}V^{j\rangle}$ (assuming that the origin of
the coordinates is at the initial position of the center of symmetry of
the BSS). Therefore, any contribution to $\mathrm{I}_{ij}$ coming from
the gauge parameter $C_2$ must contain, at least, the factors $C_2$ and
$V^{\langle i}V^{j\rangle}$, and also the total mass $M$. Taking into
account the dimensionality of $C_2 \sim (G M/c^2)^2 \,a$, which is that
of a length cubed, it is easily seen that there is no way to generate
such a contribution to $\mathrm{I}_{ij}$. Therefore, we conclude that
the source quadrupole moment of a boosted general, harmonic-coordinate
spherically symmetric metric is strictly equal to the source quadrupole
moment of a boosted \textit{standard harmonic-coordinate} Schwarzschild
solution, obtained by setting $C_2 =0$ (and $C_1 =1$) in~(\ref{gensol}),
\textit{i.e.} by choosing the standard harmonic radial
coordinate~(\ref{standsol}).

In the following, we shall therefore consider only such a boosted
Schwarzschild solution (BSS) in standard form. We shall sometimes refer
to the source of this solution as a black hole (though, strictly
speaking, one should always have in mind some extended spherical star).
For simplicity, we shall translate the origin of the coordinate system
so that it is located at the initial position of the black hole at
coordinate time $t=0$. With this choice of origin of the coordinates all
the current-type moments $\mathrm{J}_L$ of the BSS are zero. We shall
concentrate our attention on the mass-type quadrupole moment
$\mathrm{I}_{ij}$, that we shall compute at the 3PN order.

The BSS metric, written in terms of the Gothic metric deviation
$h^{\mu\nu}=\sqrt{-g}\,g^{\mu\nu}-\eta^{\mu\nu}$, satisfying standard
harmonic coordinates so $\partial_\nu h^{\mu\nu}=0$, is best formulated
in a manifestly Lorentz covariant way as
\begin{equation}\label{BSScov}
h^{\mu\nu} = \left(1-\frac{\left(1+\frac{G\,M}{c^2\,r_\perp}\right)^3}{1
  -\frac{G\,M}{c^2\,r_\perp}}\right)u^\mu u^\nu -
\frac{G^2\,M^2}{c^4\,r_\perp^2} \,n^\mu n^\nu\,,
\end{equation}
where $u^\mu$ is the time-like unit four-velocity of the center of
symmetry of the BSS, where $n^\mu$ is the space-like unit vector
pointing from the BSS to the field point along the direction
\textit{orthogonal} (in a Minkowskian sense) to the world line of the
BSS, and where $r_\perp$ denotes the orthogonal distance to the world
line (square root of the interval). However, in our explicit
calculations (done with the software Mathematica) it is preferable to
employ a more ``coordinate-rooted'' formulation of the BSS metric, which
is of course completely equivalent to (though less elegant than)
Eq.~(\ref{BSScov}).

Le us denote by $x^\mu=(c\,t,\mathbf{x})$ the global reference frame, in
which the black hole is moving, and by $X^\mu=(c\,T,\mathbf{X})$ the
rest frame of the black hole --- both $x^\mu$ and $X^\mu$ are assumed to
be harmonic coordinates. Let $x^i(t)$ be the rectilinear and uniform
trajectory of the center of symmetry of the BSS in the global
coordinates $x^\mu$, and $\mathbf{V}=(V^i)$ be the constant coordinate
velocity of the BSS,
\begin{equation}\label{V}
V^i\equiv \frac{d x^i(t)}{d t}\,.
\end{equation}
The rest frame $X^\mu$ is transformed from the global one $x^\mu$ by the
Lorentz boost,
\begin{equation}\label{xboost}
x^\mu = \Lambda^\mu_{~\nu}(\mathbf{V})\,X^\nu\,.
\end{equation}
For simplicity we consider a pure Lorentz boost
$\Lambda^\mu_{~\nu}(\mathbf{V})$ without rotation of the spatial
coordinates. As explained above, we can assume that the metric of the
BSS in the rest frame $X^\mu$ takes the standard harmonic-coordinate
Schwarzschild expression, which we write in terms of the Gothic metric
deviation $H^{\mu\nu}$, satisfying $\partial_\nu H^{\mu\nu}=0$. Hence,
\begin{eqnarray}
H^{00} &=&
1-\frac{\left(1+\frac{G\,M}{c^2\,R}\right)^3}{1-\frac{G\,M}{c^2\,R}}\,,\\
H^{i0} &=& 0\,,\\ H^{ij} &=&
-\frac{G^2\,M^2}{c^4\,R^2}\,N^iN^j\,,\label{Hij}
\end{eqnarray}
where $M$ is the total mass, $R\equiv\vert\mathbf{X}\vert$ and
$N^i\equiv X^i/R$. A well-known feature of the Schwarzschild metric in
harmonic coordinates is that the spatial Gothic metric $H^{ij}$ is made
of a single quadratic-order term $\propto G^2$ as shown in
Eq.~(\ref{Hij}). The Gothic metric deviation transforms like a Lorentz
tensor so the metric of the BSS in the global frame $x^\mu$ reads as
\begin{equation}\label{hmunu}
h^{\mu\nu}(x) =
\Lambda^\mu_{~~\rho}\Lambda^\nu_{~~\sigma}\,H^{\rho\sigma}(\Lambda^{-1}x)\,,
\end{equation}
in which the rest-frame coordinates have been expressed by means of the
global ones, \textit{i.e.} $X^\mu(x)=(\Lambda^{-1})^\mu_{~\nu} x^\nu$,
using the inverse Lorentz transformation. The only problem is to derive
the explicit relations giving the rest-frame radial coordinate $R$ and
the unit direction $N^i$ as functions of their global-frame counterparts
$r$ and $n^i$, of the global coordinate time $t$, and of the boost
velocity $V^i$. For these relations we find
\begin{eqnarray}
R&=&r\,\biggl[1+c^2(\gamma^2-1)\left(\frac{t}{r}\right)^2-2\gamma^2
(Vn)\left(\frac{t}{r}\right)+\gamma^2\frac{(Vn)^2}{c^2}\biggr]^{1/2}\,,\label{R}\\
N^i &=& \frac{r}{R} \left[n^i-\gamma
V^i\left(\frac{t}{r}\right)+\frac{\gamma^2}{\gamma+1}\frac{V^i}{c^2}(Vn)\right]\,,\label{Ni}
\end{eqnarray}
where $\gamma\equiv\left(1-V^2/c^2\right)^{-1/2}$ and $(Vn)\equiv
\mathbf{V}\cdot\mathbf{n} = V^jn^j$ is the usual Euclidean scalar
product. The latter formulation of the BSS metric is well adapted to our
calculations because we shall have to perform, for computing the source
multipole moments, an integration over the \textit{coordinate}
three-dimensional spatial slice $\mathbf{x}\in\mathbb{R}^3$, with
\textit{coordinate} time $t=\mathrm{const}$, which is easily done using
the explicit relations~(\ref{R})--(\ref{Ni}).

\subsection{Quadrupole moment of a boosted Schwarzschild black hole}\label{secVB}

We compute the quadrupole moment $\mathrm{I}_{ij}$ of the BSS, following
the prescriptions defined by Eq.~(\ref{ILfinal}). To this end we first
expand $h^{\mu\nu}$ when $c\rightarrow +\infty$, taking into account all
the $c$'s present both in the expression of the rest frame metric
$H^{\mu\nu}$ as well as those coming from the Lorentz
transformation~(\ref{hmunu})--(\ref{Ni}). In this process the boost
velocity $\mathbf{V}$ is to be considered as a constant, ``spectator'',
vector. Note in passing that, in the present problem, the characteristic
size $a$ of the source at time $t$ is given by the displacement from the
origin, $a \sim V \,t$, where $V\equiv \vert\mathbf{V}\vert$, while the
near-zone corresponds to $r \ll c\,t$. Therefore, the far near-zone,
where we read off the multipole moments as some combination of expansion
coefficients $\varphi_{a,b}^{\mu\nu}(\mathbf{n},t)$, is the domain $V
\,t \ll r \ll c\,t$. We have evidently to assume that $V\ll c$ for this
region to exist.

We then first get the near-zone (or PN) expansion of the BSS metric,
$\overline{h}^{\mu\nu}$, by expanding in inverse powers of $c$ up to 3PN
order. Next we compute the multipolar (or far) re-expansion of each of
the PN coefficients when $r\rightarrow +\infty$ with $t=\mathrm{const}$.
In this way we obtain what we have denoted by
$\mathcal{M}\bigl(\overline{h}^{\mu\nu}\bigr)$ in Eq.~(\ref{Mhexp}). In
the BBS case it is evident that the far-zone expansion given
by~(\ref{Mhexp}) involves simply some powers of $1/r$, without any
logarithm of $r$.

With $\mathcal{M}\bigl(\overline{h}^{\mu\nu}\bigr)$ in hand we have the
coefficients of the various powers of $1/r$, and we obtain thereby the
needed quantities $\Psi_{k,\ell}^{\mu\nu}$ defined by Eq.~(\ref{Psi}).
It is then a simple matter to compute all the required angular averages
present in the formula~(\ref{ILfinal}) and to obtain the following 3PN
mass quadrupole moment of the BSS,
\begin{eqnarray}\label{IijBSS}
\mathrm{I}_{ij}^\mathrm{\,BSS} &=& M\,t^2 \,V^{\langle
i}V^{j\rangle}\left[1+\frac{9}{14}\,\frac{V^2}{c^2}
+\frac{83}{168}\,\frac{V^4}{c^4}
+\frac{507}{1232}\,\frac{V^6}{c^6}\right] \nonumber\\ &+&
\frac{4}{7}\,\frac{G^2\,M^3}{c^6}\,V^{\langle i}V^{j\rangle} +
\mathcal{O}\left(\frac{1}{c^8}\right)\,.
\end{eqnarray} 
The first term represents the standard Newtonian expression, augmented
here by a bunch of relativistic corrections. (Recall that we have chosen
the origin of the coordinate system at the initial location of the BSS
at $t=0$.)

The last term in Eq.~(\ref{IijBSS}), with coefficient $\mathcal{C}=4/7$,
is the most interesting for our purpose. It is purely of 3PN order, and
it contains the first occurence of the gravitational constant $G$, which
therefore arises in the quadrupole of the BSS only at 3PN order. This
term is interesting because it corresponds to one of the regularization
ambiguities, due to an incompleteness of Hadamard's self-field
regularization, which appears in the calculation of the mass-type
quadrupole moment of point particle binaries at the 3PN
order~\cite{BIJ02,BI04mult}. As we see from Eq.~(\ref{Iij}) the
associated ambiguity parameter is $\zeta$, which represents in fact the
analogue of the kinetic ambiguity parameter $\omega_\mathrm{kinetic}$ in
the equations of motion, see Eq.~(\ref{omkin}). It is now clear that
$\zeta$ can be determined from what we shall now call the \textit{BSS
limit} of a binary system, which consists of setting one of the masses
of the binary to be \textit{exactly zero}, say $m_2=0$.

We have obtained the BSS limit of the 3PN mass-type quadrupole moment of
compact binaries computed for general binary orbits in
Refs.~\cite{BIJ02,BI04mult}. We have also inserted for the position of
the first body $y_1^i=v_1^i\,t$ in order to conform with our choice for
the origin of the coordinates. In this way we get
\begin{eqnarray}\label{Iijbin}
\mathrm{I}_{ij}^\mathrm{\,BSS\,limit} &=& m_1\,t^2 \,v_1^{\langle
i}v_1^{j\rangle}\left[1+\frac{9}{14}
\,\frac{v_1^2}{c^2}+\frac{83}{168}\,\frac{v_1^4}{c^4}
+\frac{507}{1232}\,\frac{v_1^6}{c^6}\right]\nonumber\\ &+&
\biggl(\frac{232}{63}+\frac{44}{3}\zeta\biggr)
\frac{G^2\,m_1^3}{c^6}\,v_1^{\langle i}v_1^{j\rangle} +
\mathcal{O}\left(\frac{1}{c^8}\right)\,.
\end{eqnarray}
The comparison of Eqs.~(\ref{Iijbin}) and~(\ref{IijBSS}) reveals a
complete match between the two results if and only if we have the
expected agreement between the masses, \textit{i.e.} $M=m_1$, and the
velocities, $v_1^i=V^i$ (since the velocity of the body remaining after
taking the BSS limit should exactly be the boost velocity), and the
ambiguity constant $\zeta$ takes the \textit{unique} value
\begin{equation}\label{zetares}
\zeta = -\frac{7}{33}\,.
\end{equation} 
Our conclusion, therefore, is that the ambiguity parameter $\zeta$ is
uniquely determined by the BSS limit. Because of the close relation
between the BSS limit with Lorentz boosts, it is clear that $\zeta$ is
linked to the Lorentz-Poincar\'e invariance of the multipole moment
formalism of Ref.~\cite{B98mult} as applied to compact binary systems
in~\cite{BIJ02,BI04mult}. This link strongly suggests that the specific
value~(\ref{zetares}) represents the only one for which the expression
of the 3PN quadrupole moment is compatible with the Poincar\'e symmetry.
In other words the present calculation indicates that the Poincar\'e
invariance should correctly be incorporated into the laws of
transformation of the source-type multipole moments for general extended
PN sources as given by Eqs.~(\ref{IL})--(\ref{JL})
or~(\ref{ILfinal})--(\ref{JLfinal}).

Let us finally emphasize that Eq.~(\ref{zetares}) has been obtained here
without using any regularization scheme for curing the divergencies
associated with the self field of point particles. However, we find,
very nicely, that the value for $\zeta$ is in agreement with the one
derived in the problem of point particles binaries at 3PN order by means
of the dimensional self-field regularization, Eq.~(\ref{zeta}). This
shows in particular that dimensional regularization is able to correctly
keep track of the global Poincar\'e invariance of the general
relativistic description of isolated systems.

\bibliography{/home/blanchet/Articles/ListeRef/ListeRef}

\begin{thebibliography}{10}

\bibitem{3mn}
C.~Cutler, T.A. Apostolatos, L.~Bildsten, L.S. Finn, E.E. Flanagan,
  D.~Kennefick, D.M. Markovic, A.~Ori, E.~Poisson, G.J. Sussman, and K.S.
  Thorne.
\newblock The last three minutes: Issues in gravitational-wave measurements of
  coalescing compact binaries.
\newblock {\em Phys. Rev. Lett.}, 70:2984--2987, 1993.

\bibitem{CF94}
C.~Cutler and E.E. Flanagan.
\newblock Gravitational waves from merging compact binaries: How accurately can
  one extract the binary's parameters from the inspiral waveform?
\newblock {\em Phys. Rev. D}, 49:2658--2697, 1994.

\bibitem{BDIWW95}
Luc Blanchet, Thibault Damour, Bala~R. Iyer, Clifford~M. Will, and Alan.~G.
  Wiseman.
\newblock Gravitational radiation damping of compact binary systems to second
  post-newtonian order.
\newblock {\em Phys. Rev. Lett.}, 74:3515--3518, 1995.

\bibitem{BFIJ02}
Luc Blanchet, Guillaume Faye, Bala~R. Iyer, and Benoit Joguet.
\newblock Gravitational-wave inspiral of compact binary systems to 7/2
  post-newtonian order.
\newblock {\em Phys. Rev. D}, 65:061501(R), 2002.
\newblock Erratum Phys. Rev. D {\bf 71}, 129902(E) (2005).

\bibitem{BDEI04}
Luc Blanchet, Thibault Damour, Gilles Esposito-Far{\`e}se, and Bala~R. Iyer.
\newblock Gravitational radiation from inspiralling compact binaries completed
  at the third post-newtonian order.
\newblock {\em Phys. Rev. Lett.}, 93:091101, 2004.

\bibitem{JaraS98}
P.~Jaranowski and G.~Sch\"afer.
\newblock Third post-newtonian higher order adm hamilton dynamics for two-body
  point-mass systems.
\newblock {\em Phys. Rev. D}, 57:7274--7291, 1998.

\bibitem{JaraS99}
P.~Jaranowski and G.~Sch\"afer.
\newblock Binary black-hole problem at the third post-newtonian approximation
  in the orbital motion: Static part.
\newblock {\em Phys. Rev. D}, 60:124003--1--12403--7, 1999.

\bibitem{BF00}
Luc Blanchet and Guillaume Faye.
\newblock Equations of motion of point-particle binaries at the third
  post-newtonian order.
\newblock {\em Phys. Lett. A}, 271:58, 2000.

\bibitem{BFeom}
Luc Blanchet and Guillaume Faye.
\newblock General relativistic dynamics of compact binaries at the third
  post-newtonian order.
\newblock {\em Phys. Rev. D}, 63:062005, 2001.

\bibitem{ABF01}
V.C. de~Andrade, L.~Blanchet, and G.~Faye.
\newblock Third post-newtonian dynamics of compact binaries: Noetherian
  conserved quantities and equivalence between the harmonic-coordinate and
  adm-hamiltonian formalisms.
\newblock {\em Class. Quant. Grav.}, 18:753--778, 2001.

\bibitem{B98tail}
Luc Blanchet.
\newblock Gravitational-wave tails of tails.
\newblock {\em Class. Quant. Grav.}, 15:113--141, 1998.
\newblock Erratum Class. Quant. Grav. {\bf 22}, 3381 (2005).

\bibitem{BIJ02}
Luc Blanchet, Bala~R. Iyer, and Benoit Joguet.
\newblock Gravitational waves from inspiralling compact binaries: Energy flux
  to third post-newtonian order.
\newblock {\em Phys. Rev. D}, 65:064005, 2002.
\newblock Erratum Phys. Rev. D {\bf 71}, 129903(E) (2005).

\bibitem{BI04mult}
Luc Blanchet and Bala~R. Iyer.
\newblock Hadamard regularization of the third post-newtonian gravitational
  wave generation of two point masses.
\newblock {\em Phys. Rev. D}, 71:024004, 2004.

\bibitem{Hadamard}
J.~Hadamard.
\newblock {\em Le probl\`eme de Cauchy et les \'equations aux d\'eriv\'ees
  partielles lin\'eaires hyperboliques}.
\newblock Hermann, Paris, 1932.

\bibitem{Schwartz}
L.~Schwartz.
\newblock {\em Th\'eorie des distributions}.
\newblock Hermann, Paris, 1978.

\bibitem{tHooft}
G.~'t~Hooft and M.~Veltman.
\newblock {\em Nucl. Phys.}, B44:139, 1972.

\bibitem{DJSdim}
T.~Damour, P.~Jaranowski, and G.~Sch\"afer.
\newblock Dimensional regularization of the gravitational interaction of point
  masses.
\newblock {\em Phys. Lett. B}, 513:147--155, 2001.

\bibitem{BDE04}
Luc Blanchet, Thibault Damour, and Gilles Esposito-Far{\`e}se.
\newblock Dimensional regularization of the third post-newtonian dynamics of
  point particles in harmonic coordinates.
\newblock {\em Phys. Rev. D}, 69:124007, 2004.

\bibitem{BDEI05dr}
Luc Blanchet, Thibault Damour, Gilles Esposito-Far{\`e}se, and Bala~R. Iyer.
\newblock Dimensional regularization of the third post-newtonian gravitational
  wave generation of two point masses.
\newblock {\em Phys. Rev. D}, 71:124004, 2005.

\bibitem{BFreg}
Luc Blanchet and Guillaume Faye.
\newblock Hadamard regularization.
\newblock {\em J. Math. Phys.}, 41:7675--7714, 2000.

\bibitem{BFregM}
Luc Blanchet and Guillaume Faye.
\newblock Lorentzian regularization and the problem of point-like particles in
  general relativity.
\newblock {\em J. Math. Phys.}, 42:4391--4418, 2001.

\bibitem{D83houches}
T.~Damour.
\newblock Gravitational radiation and the motion of compact bodies.
\newblock In N.~Deruelle and T.~Piran, editors, {\em Gravitational Radiation},
  pages 59--144, Amsterdam, 1983. North-Holland Company.

\bibitem{Kop85}
S.M. Kopeikin.
\newblock The equations of motion of extended bodies in general-relativity with
  conservative corrections and radiation damping taken into account.
\newblock {\em Astron. Zh.}, 62:889--904, 1985.

\bibitem{GKop86}
L.P. Grishchuk and S.M. Kopeikin.
\newblock Equations of motion for isolated bodies with relativistic corrections
  including the radiation reaction force.
\newblock In J.~Kovalevsky and V.A. Brumberg, editors, {\em Relativity in
  Celestial Mechanics and Astrometry}, pages 19--33, Dordrecht, 1986. Reidel.

\bibitem{itoh1}
Yousuke Itoh and Toshifumi Futamase.
\newblock {\em Phys. Rev. D}, 68:121501(R), 2003.

\bibitem{itoh2}
Yousuke Itoh.
\newblock {\em Phys. Rev. D}, 69:064018, 2004.

\bibitem{EIH}
A.~Einstein, L.~Infeld, and B.~Hoffmann.
\newblock The gravitational equations and the problem of motion.
\newblock {\em Ann. Math.}, 39:65--100, 1938.

\bibitem{B98mult}
Luc Blanchet.
\newblock On the multipole expansion of the gravitational field.
\newblock {\em Class. Quant. Grav.}, 15:1971--1999, 1998.

\bibitem{BDI04zeta}
Luc Blanchet, Thibault Damour, and Bala~R. Iyer.
\newblock Surface-integral expressions for the multipole moments of
  post-newtonian sources and the boosted schwarzschild solution.
\newblock {\em Class. Quant. Grav.}, 22:155, 2005.

\bibitem{Dgef96}
Thibault Damour and Gilles Esposito-Far{\`e}se.
\newblock Testing gravity to second post-newtonian order: A field theory
  approach.
\newblock {\em Phys. Rev. D}, 53:5541--5578, 1996.

\bibitem{Weinberg}
S.~Weinberg.
\newblock {\em Gravitation and Cosmology}.
\newblock John Wiley, New York, 1972.

\bibitem{Fock}
V.A. Fock.
\newblock {\em Theory of space, time and gravitation}.
\newblock Pergamon, London, 1959.

\end{thebibliography}

%
%

\end{document}